# Mathematical model of fluid flow in an osteon Influence of cardiac system

W. MILADI*† and M. RACILA‡

† [1]Department of Applied Mechanics, FEMTO ST Institute, Besançon, France
walidmiladi@hotmail.com

‡ [1]Department of Applied Mathematics, University of Craiova, Romania
mracila@yahoo.com

*Corresponding author. Email: walidmiladi@hotmail.com

## 1 Introduction

Numerical simulations of the behavior of the osteonal structure are more and more acute and an important parameter is the pressure of the bony fluid. Haversian and Volkmann canals contain blood vessels that transport oxygen and nutrients necessary for the cellular activity. The pressure in these vessels must be taken into account. While it is possible to estimate the value of this pressure, there is no information on the effect of the vessel wall that may have in the transmission of pressure. This point is important for the applications that can be done with the SiNuPrOs model [1, 2] since the pressure variations are very low in such a structure (because its small size). Moreover, taking into account the mesh of the wall in the simulations increases the size of the problem and the cost of its solving.

## 2 Methods

We consider a femoral bone for which one knows the parameters of its architecture. The SiNuPrOs model allows computing firstly the macroscopic homogenized elastic properties and then the elastic properties of an osteon and of surrounding interstitial system. This study is based on the localization aspect of the SiNuPrOs model. Knowing a macroscopic loading on this bone, it's possible to compute, by a FEM, all the mechanical fields at the macroscopic scale and to determine these fields at the scale of one osteon.

The osteon {O} is considered as a cylinder embedded in a parallelepiped box of interstitial system {IS} and is crossed by the haversian channel {H} where a cylindrical hole represents a blood vessel (Figure 1). The flow is modeled by a Brinkmann equation in {H}, a conservation equation with the Darcy's law in {O} U {IS} :

$$-\operatorname{div}(-pId + \mu\Delta v) + \frac{\mu}{\kappa}v = G \quad \text{in } \{H\}$$

$$-\operatorname{div}[\frac{\kappa}{\mu\varepsilon}(\nabla P + \rho g)] = 0 \quad \text{in } \{O\} \cup \{IS\}$$

where $p$, $v$, $\kappa$, $\mu$, $\varepsilon$ and $G$ are respectively the pressure, the velocity, the permeability, the viscosity, the porosity and the gravitational force.

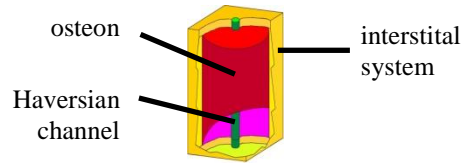

Figure 1. Osteonal structure

Two cases are studied: in the first one has only these domains and in the second, a thin layer is introduced inside the haversian channel to model the wall of the vessel.

Some conditions of coupling are added. On the interface between {H} and in {O} U {IS} one considers an equality of pressure and a correlation depending of the porosity between the velocities. On the interface between the vessel wall and {H}, one imposes equality between the normal component of the stresses and the fluid pressure and equalities between the fluid velocity and the velocity of the external boundary of the wall. A periodic function is applied on the fluid inside H in order to simulate the blood pressure for which the minimal and maximal values are respectively 9.69 and 9.71Kpa (Figure 2).

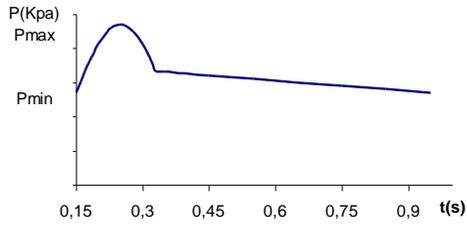

Figure 2. Evolution of the blood pressure during a cardiac cycle

## 3 Results and Discussion

On a mathematical point of view, it is possible to prove that the two previous problems have unique solutions in velocity and pressure. In order to evaluate the effect of the wall, we study the evolution of the pressure in several points located inside the osteon but not in the immediate vicinity of the haversian channel in the framework of a comparative analysis between the cases with and without wall effect. The considered parameters are the fluid viscosity and the elastic properties of the wall. The fluid viscosity varies between 1 and 4 mPa.s, the Young's modulus between 0.1 and 10 MPa and the presented results (Figure 3) are those corresponding to the maximal value of the blood pressure.

The only difference that we can observe in each point between the two cases only deals with the intensities of the considered physical quantities. There in no major differences between the various points.

We find that the wall vessel absorbs a part of the blood pressure and for a Young modulus higher than 0.5 MPa, the absorption is almost constant for each viscosity.

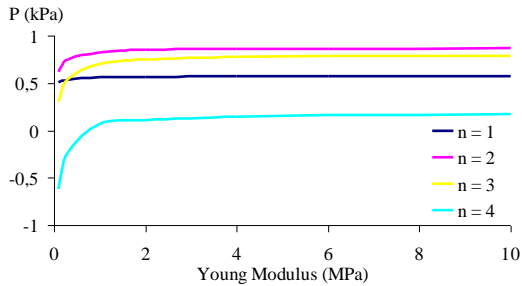

Figure 3. Decreasing of the pressure induced by the vessel wall (stationnary case)

One can see on Figure 3 the similarity of the curves. It should be noted that the presence of the wall induces a decreasing of the pressure which can be important (until 0.8 kPa). This effect cannot be neglected since the mean value of the order of magnitude of the pressure in the osteonal structure is 1-10 kPa. It seems that this influence is constant for every Young's modulus that is higher than 1 MPa. In fact, it is not exactly constant and since the local variations in this osteonal structure are low (about 10-80 Pa), this variations can have some importance. These variations for a Young's modulus of the vessel wall varying between 1 et 10 MPa are summarized on Table 1.

| Viscosity (mPa.s) | 1 | 2 | 3 | 4 |
|---|---|---|---|---|
| Pressure variation (Pa) | 10 | 39 | 82 | 100 |

Table 1. Variation of the pressure for a Young's modulus of the vessel wall varying between 1 et 10 MPa as function of the fluid viscosity.

The last part of this study concerns the effect of the pulsatile blood flow. This pulsatile aspect is conserved in the osteonal structure but there is a smoothing of the values and a decreasing of the maximal value due to the low permeabilities of the osteonal and interstitial system structures when the mineralization degree is important.

## 4 Conclusions

This study shows that good quantification of pressure in the osteonal structure requires the taking into account of blood vessels in the channels of Havers and Volkmann and the introduction of an amplitude pressure to find the effect of pulsatile blood flow.

There is no need to mesh the wall, which would be very expensive and it is only necessary to change the value of the applied pressure.

The multi scale analysis developed by SiNuPrOs allows various investigations using datasets consistent with each other.

## Acknowledgments

Grant PN-II-RU-RP-2008-4-7 of the Romanian MEC
We thank J.M. Crolet for his remarks.